# HEP Applications Evaluation of the EDG Testbed and Middleware

I. Augustin, F. Carminati, J. Closier, E. van Herwijnen
*CERN, Geneva, Switzerland*

J. J. Blaising, D. Boutigny
*LAPP-Annecy, CNRS, France*

C. Charlot
*Ecol.Poly, Paris, France*

V. Garonne, A. Tsaregorodtsev
*Université d'Aix - Marseille II, CNRS, France*

K. Bos, J. Templon
*NIKHEF, Amsterdam, Holland*

P. Capiluppi, A. Fanfani
*Univ. Bologna, INFN, Italy*

R. Barbera
*Univ. Catania, INFN, Italy*

G. Negri, L. Perini, S. Resconi
*Univ. Milan, INFN, Italy*

M. Sitta
*Univ. Piemonte, INFN, Italy*

M. Reale
*Univ. Rome "La Sapienza", INFN, Italy*

D. Vicinanza
*Univ. Salerno, INFN, Italy*

S. Bagnasco, P. Cerello
*Univ. Turin, INFN, Italy*

A. Sciaba
*CNAF/Bologna, INFN, Italy*

O. Smirnova
*Lund University, Lund, Sweden*

D. Colling
*Imperial College, Univ. London, UK*

F. Harris
*Particle Physics, Univ. Oxford, UK*

S. Burke
*Rutherford Appleton Laboratory, Oxon, UK*

Workpackage 8 of the European Datagrid project was formed in January 2001 with representatives from the four LHC experiments, and with experiment independent people from five of the six main EDG partners. In September 2002 WP8 was strengthened by the addition of effort from BaBar and D0. The original mandate of WP8 was, following the definition of short- and long-term requirements, to port experiment software to the EDG middleware and testbed environment. A major additional activity has been testing the basic functionality and performance of this environment. This paper reviews experiences and evaluations in the areas of job submission, data management, mass storage handling, information systems and monitoring. It also comments on the problems of remote debugging, the portability of code, and scaling problems with increasing numbers of jobs, sites and nodes. Reference is made to the pioneeering work of Atlas and CMS in integrating the use of the EDG Testbed into their data challenges. A forward look is made to essential software developments within EDG and to the necessary cooperation between EDG and LCG for the LCG prototype due in mid 2003.

## 1. INTRODUCTION

In the last year the LHC experiments, in the context of Workpackage 8 (WP8) of the European Datagrid project (EDG), have performed extensive tests with the EDG testbed. In addition BaBar and D0 have accomplished some preliminary evaluations. This paper summarises the experiences so far.

Many aspects of EDG middleware and the experiences of specific HEP experiments with various grid projects are reported elsewhere at this conference [3-14], hence this paper gives only a brief resumé in some areas.

## 2. THE EDG PROJECT

### 2.1. Project Structure

The European Datagrid is a three-year project which started in January 2001, funded by the EU and six main partner organisations (CERN, CNRS, ESA, INFN, NIKHEF, and PPARC). The aim is to develop a fully-fledged, operating grid. The work is split into units known as workpackages (WPs). There are six WPs which develop grid middleware, in the areas of job submission, data management, information and monitoring, computing fabric management, mass storage and networking, plus a cross-cutting Security Co-ordination Group which considers authentication and authorisation issues. There is a Testbed WP which deals with the configuration and management of the various EDG testbed machines. Finally, there are three application WPs, in the areas of High Energy Physics, Earth Observation and Biomedical applications.

The latest production release of the EDG software is version 1.4, and this is the version evaluated in this paper. A major new version 2.0 is expected in May, with many new features, and some indications are given of where improvements are expected.

### 2.2. WP8

WP8 has two main functions; on one hand to capture the requirements of HEP experiments and transmit them to the middleware developers, and on the other hand to pass on knowledge about the testbed to the experiments to enable them to use it for their Data Challenges. In addition there has been a substantial effort invested in general testing and debugging of the middleware. A report has to be sent to the EU each year giving the evaluation of the current release. The latest report [1] is the basis for this paper.

WP8 members are also involved in various other activities, e.g. architecture, quality control, user tutorials, user support etc.

The WP8 membership consists of a full-time manager and five full-time Experiment Independent Persons (EIPs). In addition each experiment contributes 2-3 representatives to the main technical working group (TWG).

## 3. THE CURRENT EDG SYSTEM

### 3.1. The Application Testbed

EDG has a number of testbeds [4] for various purposes, but users largely interact with the application testbed. The nature of the project means that this is not operated fully as a production system, but in general the intention is for it to run in a stable way with software that has been through some basic testing.

The testbed has run essentially continuously since November 2001. For much of that time it consisted of machines at the five "core sites" related to the main partner organisations: CERN, CNAF, IN2P3, NIKHEF and RAL. It is now expanding rapidly; in March 2003 there were about 15 sites in the system (including sites in Asia and the US) with around 900 CPUs, 10 Tb of permanent disk storage and four sites with access to tape-based mass storage.

The evolution of the testbed is described in more detail elsewhere [4], but some key dates were as follows. The first fully functional release was version 1.1 in February 2002. This was followed by version 1.2 in April 2002, which was the first version with sufficient stability to be used for serious work. Atlas pioneered this in August 2002. Atlas found some serious problems, which resulted in the 1.3 release in November 2002 (incorporating an upgrade to the Globus version) which was used by CMS for the beginning of their stress test. Evolution has continued in December 2002 and 2003 with the 1.4.x series of releases, which largely contain fixes for bugs found by users (now also including Alice and LHCb) and has much improved stability. A major new version 2.0 is expected for May 2003, which will incorporate substantial new software in many areas, and will form the basis of both the final EDG testbed and the initial production testbed for the LCG project.

### 3.2. EDG Middleware

The middleware in the EDG 1.x releases is based on various core Globus services: GridFTP, MDS, the Globus Replica Catalog and the Globus gatekeeper/job manager. EDG has added new middleware in four main areas, and a brief summary is given here.

**Job submission**: A job description in the Condor *classads* language is matched against information in the MDS, and the job is dispatched to the best-matching site. Job information is kept in a logging and bookkeeping database [6].

**Data management**: tools are provided to replicate files using GridFTP, and register them in the Replica Catalogs. There are also hooks to stage files between disk storage and a tape-based Mass Storage System. The data management tools have an interface to the job submission system to allow jobs to be steered to their input files [5].

**Fabric management**: there is an extensive system to install, configure and manage large farms [13].

**VO management**: tools are provided to register users within a VO, and to map users to dynamically-allocated local accounts at each site [14].

## 4. WP8 ACTIVITIES

### 4.1. Data Challenges

Atlas and CMS performed extensive exercises in 2002, while Alice and LHCb started using the testbed in February 2003. Other papers describe their activities in detail, so only a brief summary is given here.

#### 4.1.1. Atlas

Atlas made the first attempt at serious use of the testbed, starting in August 2002. The intention was to repeat part of the recently-completed Monte Carlo Data Challenge [12]. A taskforce was formed consisting of people from the Atlas experiment, from the EDG middleware groups, and the WP8 EIPs. The tests were broadly successful, but identified two major problems which were fixed in the 1.3 release in November 2002.

#### 4.1.2. CMS

CMS also followed the task force approach, but with a more ambitious test aiming to perform part of their real Data Challenge. Over three weeks they succeeded in simulating about 250,000 events [9], using an extended testbed which added some CMS-dedicated resources.

#### 4.1.3. ALICE

In March 2003 ALICE [11] started a production of 5,000 simulated heavy-ion events, which should amount to approximately 9 Tb of data, taking around 120k CPU-hours. In this case the EDG testbed is integrated into the existing ALICE system as a single large computing resource.

#### 4.1.4. LHCb

LHCb [10] has also been using the testbed since February 2003 and has simulated approximately 200,000 events as part of a Physics Production Data Challenge, with a similar system architectural approach to ALICE.

#### 4.1.5. BaBar and D0

These experiments are already taking data, and hence are waiting for a production service before making serious use of the testbed, but they have both performed limited tests [8].

### 4.2. Use Cases

As described elsewhere [3], in May 2002 a document [2] was produced to collect common Use Cases for the four LHC experiments. This exercise was initially started in WP8 and then continued in the LCG framework. WP8 has recently assessed the extent to which release 1.4 of the EDG middleware can support the Use Cases [1].

Of the 43 Use Cases, 6 are fully implemented, and a further 12 are largely satisfied but have some restrictions or complications. 16 are not implemented because essential functionality is missing. The remaining 9 are partially implemented, but have significant features which are missing in the current version.

The missing features fall into three general categories. Some Use Cases relate to Virtual Data, which is outside the scope of the EDG project. A second category relate to areas like authorisation, job control and optimisation, where we expect significant improvements in release 2.0.

The third category relates to the question of metadata catalogues. There is some support from the middleware in this area, but the experiments will need to clarify their needs before it becomes clear if the support is sufficient.

## 5. LESSONS FOR THE FUTURE

In an experimental system it is inevitable that problems will be found, and the real test of success is the speed with which bugs are fixed and problems are solved. WP8 has had an excellent working relationship with the middleware developers, testbed managers and integration team, and we have made good progress towards a stable system which can be used in a real production environment. This section collects some comments and experiences with the current system, many of which may have a broader relevance than just the EDG project. We also indicate where improvements are expected in the next major release of the middleware.

### 5.1. General

One of the most significant lessons from EDG is the importance of having a large, operating testbed, run as a quasi-production system. Many problems have been found which were not seen in local testing by the developers, and further problems only emerged once Atlas and then CMS started trying to use the system to do real work on a large scale.

Related to this is the fact that problems of configuring and integrating software to build a complete system are at least as important as bugs in the software itself. EDG has a large number of software components, often with many configuration options, and the interactions between different systems mean that many apparently innocuous changes to the configuration can produce unexpected consequences. Grid projects should expect to devote substantial time and manpower to integration and configuration.

Grids imply some changes in the way users and system managers think about systems. In general, users are used to using a few large systems under unified control, which the system managers can configure to suit the local user needs. In a grid you have a large number of systems with no unified management, which may differ widely. To achieve the goal that jobs can run on whichever machine they land the jobs must make as few demands on the local environment as possible, and the demands which are made must be capable of being advertised in the information system. Equally, system managers need to think about the effect any changes they make will have on the operation of the grid in general.

### 5.2. Job Submission

The current testbed has some restrictions [6] which come largely from the Globus and Condor versions used. The main limits are that a Resource Broker can only have a maximum of 512 active jobs, and can support roughly 20 concurrent users and submission rates of about 1000 jobs per hour. These limits are not a major problem, at least while most large-scale job submission is managed by a small number of people. Multiple brokers can be used if a higher submission rate is needed.

The matchmaking between jobs and sites uses the information published in the information system, and hence can be vulnerable to stale or incorrect information. This may mean that the system fails to find the requested resources, and the algorithm used to rank matching sites in a preference order can fail in a variety of ways. In the worst cases a single badly-configured site may act as a "black hole", attracting all jobs.

The current job submission chain is complex, involving components from several sources (EDG, Condor, Globus and the underlying batch system). There are many places where problems can arise, and tracing the exact reason for failures can be a long and difficult task even for experts. Also much of the functionality of the underlying batch system is not made available to the end-user.

The current system allows only single jobs to be submitted. The current release has no support for extensions like job dependencies (DAGs), automatic splitting or checkpointing.

### 5.3. Local Environment

The present testbed offers rather little information or control over the local environment seen by a job running on a batch worker node, for example scratch space on local or NFS-mounted disks, installed software etc. In particular there is no management of disk space, so a job cannot ensure that enough space is available to make local copies of files. Installation of experiment software has been extensively discussed, but a good working solution has not yet been reached. There is also no consensus so far on how to deal with "system-type" software which may be needed by applications, e.g. compilers, shells, *perl* modules etc.

Another problem, which is so far unsolved, is that many sites with large farms want to put the batch workers into "internet-free zones", i.e. private IP networks with no direct access to the Internet. However, both application software and the EDG middleware currently need outbound IP access from worker nodes. This is likely to become a major problem in the near future as the testbed expands.

### 5.4. Information Systems and Monitoring

The current information system uses the Globus MDS; EDG is developing an alternative system called R-GMA, but this has not yet been released. Unfortunately, EDG has not been able to configure MDS to allow it to work as designed, as a hierarchical, dynamic system. Various problems have been encountered, the most serious of which is that as the query rate and data volume increase the MDS servers slow down dramatically, taking tens of minutes to respond. This effectively paralyses the testbed. The workaround deployed in the current testbed uses a cache of the information stored in a Berkeley database LDAP back-end. However, this information is only refreshed every 15 minutes, and stale information (e.g. sites which are no longer active) is only removed infrequently by manual intervention. Also in this configuration the Resource Broker still makes direct queries to the GRIS servers running on candidate machines, and these can also exhibit the problem of very slow responses.

This situation is not really satisfactory. It is hoped that the new R-GMA system [7] in release 2.0 will improve the performance, otherwise the problems with MDS will need to be solved.

The provision of monitoring and debugging information is also quite limited at the moment, and what is available is spread among various different systems. Some general monitoring data can be derived from MDS, and this is extracted and displayed in various ways by a number of web sites. The job submission system contains its own logging and bookkeeping database, but this can currently only be queried with a command-line tool, and only for jobs submitted by the user issuing the command. Some sites have site-based monitoring, e.g. with Nagios or Ganglia. However, as grids start to make the transition to production systems there will be an increasing need for monitoring and debugging information to be available in a comprehensive and consistent form.

### 5.5. Data Management

The LDAP-based Globus Replica Catalog has not proved to be adequate for serious use. A limit on data volume translates to an effective limit of a few thousand files per file collection if the file names are of reasonable length, and there is no real support for the use of multiple catalogues. Like MDS, the catalogues respond badly to a high query rate, with queries hanging indefinitely. Also a single catalogue means a single point of failure. Release 2.0 will have a completely new catalogue system which should address these problems.

Use of the replica management tools [5] has also underlined the importance of dealing correctly with failures. Replication of large files can fail for a variety of reasons, including network problems, disk and NFS faults etc., as well as the Replica Catalog problems mentioned above, and the current system does not always leave things in a consistent state. Also there is no general consistency checking between the Catalog content and the files actually on disk, and no management of the disk space.

### 5.6. Mass Storage

Mass Storage Systems (MSS), usually based on tape robot technology, can have a variety of interfaces, often specific to a particular site. For grid use these interfaces need to be converted to a uniform system. In the current testbed this is done only to a limited extent.

EDG currently has four sites with an MSS. Sara (the Dutch national supercomputer center) has an interface which looks like a normal Unix filing system and runs a GridFTP server, and hence this can be treated like a regular disk-based Storage Element.

CERN and Lyon both use the CERN RFIO package as an interface, and RAL has a locally-written interface.

However, the EDG tools expect to use GridFTP, which is not currently available. An interim solution uses a disk-based Storage Element, with scripts being callable by the replica management tools to stage files between disk and MSS using a fixed mapping between disk file names and the location in the MSS.

This solution has been adequate so far, but has several limitations. The mapping to MSS file names is fixed, and files in the MSS cannot be directly registered in a Replica Catalog, so it is not possible to access arbitrary files in the MSS. There is no easy way to control authorisation to write to the MSS. There is also no automatic management of space on the disk used as a staging area, so users must take care not to fill the space.

There are two aspects to improving the situation. One is to add direct GridFTP access to each MSS (this is currently under test for Castor at CERN). The other is to develop middleware for management of both disk and tape-based storage; the EDG solution for this aspect will be deployed in release 2.0.

### 5.7. Virtual Organisations and Security

The core of the EDG security framework, like most grid projects, is the PKI-based Globus Security Infrastructure (GSI). Certificates are issued to users by national Certificate Authorities (CAs). This has worked well, although the time taken to approve a new CA can be rather long (several months).

Certificates are mapped to Virtual Organisations (VOs) using LDAP servers. Local tools at each site extract this information to build a grid map file, which in turn maps users into dynamically-allocated anonymous "pool" accounts, removing any need for local registration of users.

This system has generally worked well, but has a number of limitations. The VO servers are single points of failure, and this has occasionally resulted in all users being denied access at some sites. The system only allows a certificate to be mapped to a single VO, so multiple certificates are needed to join several VOs. Write access to the server uses a password so security is limited, and read access is uncontrolled.

More seriously, much of the security infrastructure is completely missing. The only authorisation control is at the level of the Unix pool accounts, and effectively there is no granularity finer than the whole VO. There are no quotas on job submission, disk usage or any other resources. Security has in many areas been an afterthought in EDG; there is no security workpackage, and different aspects of security are spread across the middleware groups, although a Security Co-ordination Group does exist to bring them together.

In general HEP does not have particularly stringent security requirements, but even so this is an important area which deserves a lot more attention. There will be significant new software in the authorisation area in release 2.0, but it remains to be seen if it will satisfy the requirements of the experiments. Also the experiments themselves will have to gain experience with VO management.

### 5.8. The Testbed

The testbed has been used by a large number of users (a few hundred), both for tests and for real production work. In the nature of the project the system has not achieved the stability needed for a real production system, but most of the time the testbed has been available to users and problems are generally fixed fairly rapidly.

Software configuration has proved to be a major problem. The grid middleware interacts in complex ways, and there is usually only one way to get things right and many ways to get them wrong. Incorrect configuration can lead to subtle failures which are hard to trace. The EDG fabric management tool (LCFG) has helped enormously because it ensures uniformity of configuration at all sites. However, this cannot be used at all sites because it needs to take complete control of the machines it manages. If grid middleware is to be installed and managed by relatively inexperienced system managers the configuration needs to become less complex and error-prone.

For the grid to become a reliable production system services need to be designed to be robust, and to fail gracefully, e.g. if resource limits are exhausted. Also in a large grid it is likely to be inevitable that individual sites will fail or will be mis-configured, and the grid as a whole needs to be protected against that as far as possible.

At present, user support is rather limited, largely based on a mailing list intended for other purposes. As the number of users increases a formal user support system will become essential.

### 5.9. Other issues

The formal requirements for document delivery by the EU mean that, unlike many software projects, EDG has an enormous amount of documentation. However, this has its own problems, and there is a need to produce condensed guides suitable for inexperienced users, and to find a more effective way to index the more detailed documentation.

At the moment most users interact with the testbed using terminal windows and command-line tools. There is also likely to be a need in the future for graphical interfaces. There are a number of projects at various levels of development, but so far there is little uniformity or convergence on common standards.

Finally, there remain the basic questions of whether the grid can be scaled to support hundreds of sites and thousands of users, and whether it will be possible to manage grids across many administrative domains. As yet the EDG testbed is not large enough to answer these questions, but the recent rapid expansion, including sites outside Europe, means that scalability will soon be tested.

## 6. THE FUTURE

EDG is currently preparing for release 2.0, expected in May 2003, which will introduce major new functionality in many areas. In particular:

- The Globus software will be taken from the VDT distribution, and will use the GLUE information schema designed in the context of the HENP Intergrid project, which should improve interoperability with the US grid projects.
- The Resource Broker has been re-designed to be more robust, and will support dependencies and checkpointing.
- R-GMA will replace MDS as the information system, with increased functionality.
- There will be a completely new system to manage mass storage systems and provide a uniform interface to them.
- The data management software will be replaced by a new system using a distributed replica catalogue.
- Job submission and data management decisions will be optimised on the basis of network performance information.

This will be the last major release of the EDG software, but we expect some further enhancements later in the year, as well as fixes to any bugs found during the summer.

The EDG project will end at the end of 2003, and the support of the testbed will progressively move into the LCG project. The first production LCG testbed is expected in July 2003, and the EDG application testbed is expected to merge with it at that time. WP8 is now co-operating with LCG in various areas (requirements gathering, testing etc.)

## 7. SUMMARY

In the first two years of the EDG project the system has gone from basic ideas to a working, multinational testbed used regularly by users from a variety of application areas. The LHC experiments, and more recently BaBar and D0, have gained a substantial amount of experience in working with grid technology. The experiments are now able to do real Monte Carlo production on the testbed.

Conversely, a great deal of feedback has been given to middleware developers, system managers and the Integration Team, with whom we have had an excellent working relationship. Members of WP8 have been involved in all areas of the project, and have provided the perspective of users to what might otherwise have been a theoretical exercise. This relationship should continue into the LCG project.

The combination of middleware development, a functioning testbed and real users has been vital to the success of the project. The involvement of running experiments (BaBar and D0, and perhaps others in the future) will also provide important information for the LHC experiments.

In the final year of the project we expect the experiments to move from a testing phase to production use of the LCG grid. We anticipate continuing the successful taskforce approach, as well as substantial generic testing of release 2.0. WP8 will also continue to support user education and architecture development activities.

## Acknowledgments

The authors wish to thank the EU and our national funding agencies for their support.

We would also like to acknowledge the active cooperation of our EDG colleagues in the middleware, networking and testbed work-packages, as well as the substantial support of the Project Office.

## References

[1] I. Augustin, F. Harris et al., Report to the EU, 'Report on results of HEP Applications at Run #1 with requirements for other WPs', April 2003
https://edms.cern.ch/document/375586

[2] F. Carminati et al., 'Common Use Cases for a HEP Common Application Layer', May, 2002
http://lcg.web.cern.ch/LCG/SC2/RTAG4/finalreport.doc

[3] F. Carminati et al., 'LHC requirements for GRID middleware', CHEP03, PSN:THCT001

[4] E. Leonardi, M. Schulz et al., 'EU Datagrid testbed management and support at CERN', CHEP03, PSN:THCT007

[5] H. Stockinger et al., 'Grid Data Management in action: Experience in running and supporting data management services in the EU Datagrid Project', CHEP03, PSN:TUAT006

[6] F. Prelz, M. Sgaravatto et al. 'The first deployment of workload management services on the Datagrid testbed: feedback on design and implementation', CHEP03, PSN:MOAT008

[7] S. Fisher et al., 'R-GMA: First results after deployment', CHEP03, PSN:MOET004

[8] D. Boutigny et al., 'Use of the European Data Grid software in the framework of the BaBar distributed computing model', CHEP03, PSN:MOCT004

[9] P. Capiluppi et al., 'Running CMS software on GRID Testbeds', CHEP03, PSN:MOCT010

[10] A. Tsaregorodtsev et al., 'DIRAC distributed implementation with remote agent control', CHEP03, PSN:TUAT006

[11] S. Bagnasco et al., 'ALICE experience with EDG', CHEP03, PSN: MOCT002

[12] G. Poulard et al., 'ATLAS Data Challenge 1', CHEP03, PSN:MOCT005

[13] O. Barring, M. Lopez et al., 'Towards automation of computing fabrics using tools from the fabric management workpackage of the EU Datagrid project', CHEP03, PSN:MODT004

[14] R. Cecchini, D. Kelsey et al., 'Managing dynamic user communities in a Grid of autonomous resources', CHEP03, PSN:TUBT005

**THCT003**